\documentclass[11pt,reqno,oneside]{amsart}
\input epsf
\usepackage{amsmath}
\usepackage{amssymb}
\usepackage{graphicx}
\usepackage{amssymb}
\usepackage{amsbsy}
\usepackage[dvips]{color}
 \newcommand{\LL}{\mathcal{L}}
\newcommand{\HH}{\mathcal{H}}
\newcommand{\tos}{\tilde o}

\begin{document}
\title[The Full Strategy Minority Game ]{The Full Strategy Minority Game }

%\subjclass{}
\author[Gabriel Acosta]{Gabriel Acosta} \author[In\'es Caridi]{In\'es Caridi}
\address{ Facultad de Ciencias Exactas y Naturales, Universidad de Buenos Aires, \\
Pabell\'on I, Ciudad Universitaria, (1428) Buenos Aires, Argentina} \email{gacosta@dm.uba.ar}
\email{ines@df.uba.ar}  
\author[ Sebasti\'an Guala]{Sebasti\'an Guala }
\author[ Javier Marenco]{Javier Marenco}
\address{ Instituto de Ciencias, UNGS, \\
J.~M.~Guti\'errez 1150, (1613) Los Polvorines, Argentina.}
\email{sguala@ungs.edu.ar }
\email{jmarenco@ungs.edu.ar }  

\thanks{}

\bibliographystyle{plain}
 
\begin{abstract}
%% Aca va el abstract
 The Full Strategy Minority Game ($FSMG$) is an  instance of the Minority Game ($MG$) which includes a single copy of every potential agent. In this work, we explicitly solve the $FSMG$ thanks to certain symmetries of this game. Furthermore, by considering the $MG$ as a statistical sample of the $FSMG$, we compute approximated values of the key variable $\sigma^2/N$ in the symmetric phase for different versions of the $MG$. As another application we prove that our results can be easily modified in order to handle certain kind of initial biased strategies scores, in particular when the bias is introduced at the agents' level. We also show that the $FSMG$ verifies a strict period two dynamics (i.e., period two dynamics satisfied with probability $1$) giving, to the best of our knowledge, the first example of an instance of the $MG$ for which this feature can be analytically proved. 
Thanks to this property, it is possible to compute in a simple way the probability that a general instance of the $MG$ breaks the period two dynamics \emph{for the first time} in a given simulation.
\end{abstract}

\maketitle
\section{Introduction}
\label{intro}

The Minority Game ($MG$) was introduced in
1997 by Challet and Zhang \cite{challet-zhang-1} in an attempt to
catch essential characteristics of a competitive population in which an individual
achieves the best result when she manages to
be in the minority group. In the $MG$, there are $N$ agents (usually odd),
that at each step of the game must choose $0$ or $1$.
Let $N_{0}(t)$ (resp.~$N_{1}(t)$) be the number of agents choosing $0$ (resp.~$1$) at the
step $t$ (note that $N_{0}(t)+N_{1}(t)=N$). The winners are those who happen
to be in the minority group (i.e., the minimum between $N_{0}(t)$ and $N_{1}(t)$). The only information
available for the agents is the system \emph{state} $\mu\in\{0,1\}^m$, that
is updated after each step of the game. The parameter $m$ is an input of the game and defines the information-processing capacity of the agents.
In the classical version of the $MG$, $\mu$ is defined to be an endogenous variable
determined from the sequence of minority sides in the last $m$ steps, although other kinds of {\it updating rules} can be found in the literature 
\cite{cavagna-irrelevance-memory,MG-per}. In early works, $\mu$ was also called  
\emph{history} and $m$  the agents' \emph{memory}.  
Therefore, the number of possible states is $\mathcal{H}=2^{m}$. Agents
play using the so-called \emph{strategies}. A strategy is a function that assigns a prediction
($0$ or $1$) for each of the possible states. In this way, there
are $\mathcal{L}=2^{\mathcal{H}}$ different strategies. Each agent has $s$
strategies at her disposal (we use $s=2$ in this work), randomly chosen with replacement from
the complete set of strategies at the beginning of the game (note that it is
possible for an agent to have two identical strategies, and for two
agents to have the same pair of strategies). At every step
of the game, each of the strategies that correctly predicted the winning side is awarded a \emph{virtual point}, regardless of use in that step. At each step, each agent plays what her best-performing strategy (in terms of
virtual points) predicts. If the two strategies have the same number of virtual points,
the agent randomly chooses one of them.

Given $N$ agents, an \emph{instance} of the $MG$ 
is a particular assignment of strategies to the agents. 
We define a \emph{configuration} $\mathcal{E}$
of the game to be a pair  $\mathcal{E}=\{\mathcal{M},I\}$, where
$\mathcal{M}= \{\tilde{\mu}^{1},\tilde{\mu}^{2},...\}$ is a sequence
of states (generated by any updating rule) and $I$ is an instance of the $MG$.

The observable $z=\ <(N_{1}-N/2)^2/N >_{\mathcal{E}}$ is the most studied and instructive  variable \cite{savit-sigma-PRL} in the $MG$. It measures the population's waste 
of resources by averaging on time and over different configurations $\mathcal{E}$, the quadratic deviation
of the number of agents that chose a fixed side (for example $N_{1}$) from $N/2$. The notation $\sigma^2/N = z$ is usual in the literature. One of the reasons why the $MG$ has attracted so much attention
is that for certain values of the parameters $m$, $N$, and $s$, the variable
$z$ is smaller than that obatined for a game 
in which each of the $N$ agents randomly chooses between
the two sides. In the latter, $N_1(t)$ corresponds to a binomial random variable $Bi(N,1/2)$, hence $z=\sigma^{2}/N =1/4$. It is interesting to note that $z$ reflects that the population as a whole achieves more resources, but does not reveal how that wealth is distributed among the agents. Ho et al.~adequatedly redefined a Gini index for the $MG$ \cite{ho-indice-gini} which showed that whenever $z$ takes its minimum value, the inequality among the agents is maximized.

For $\alpha=\mathcal{H}/N=2^{m}/N$, it was proposed  \cite{savit-sigma-PRL, challet-simetria}  that 
the curve given by $z$ against $\alpha$ is independent of $N$ (i.e., $z$ 
could be regarded as a function of the parameter $\alpha$)
although longer simulations of the $MG$ showed that the curve does depend on $N$ within the range $0.01\lesssim   \alpha\lesssim 0.2$, in the sense that the variable $z$ 
increases as $N$ increases \cite{chino-simu-largas}. However, such invariance does arise in the $MG_{rand}$ introduced by Cavagna \cite{cavagna-irrelevance-memory}, where a different updating rule is used: the system state $\mu$ is randomly chosen (with uniform distribution) at each game step. In \cite{MG-per}, the authors
propose a periodic updating rule of period $\mathcal{H}$,  which we
call $MG_{per}$ throughout this work. Like in the $MG$, $z$ increases with $N$ for the $MG_{per}$  (in the curve given by $z$ against $\alpha$) for  the same range of $\alpha$.

In the region given by $\alpha\ll1$, \emph{crowd effects} arise at some
game steps. This behavior is related to a dynamics known as \emph{Period Two Dynamics} (PTD),
which was observed for the first time by Savit et al.~\cite{manuca-ptd}. 
To understand this dynamics, that plays an important role in the rest of the article,
for each game step $t$ we define the {\it parity array} $\mathcal{P}_{\mathcal{E}}^{t}$
to be an array of
categorical variables recording the parity (odd or even) of the number of appearences of each state in the first $t-1$ game steps. More precisely, we have $\mathcal{P}_{\mathcal{E}}^{t}\in \{O,E\}^{\mathcal{H}}$ and if we identify any state $\mu$ with the integer 
number given by the binary expansion of $\mu$ plus one   (so that $\mu$ can be thought as an integer ranging from $1$ to $\HH$),  we have that   
$\mathcal{P}_{\mathcal{E}}^{t}(\mu)=O$ (resp.~$E$) if $\mu$ has appeared an odd (resp.~even) number of times in the first $t-1$ steps of the game. We assume 
$\mathcal{P}_{\mathcal{E}}^{1}=(E,E,\cdots,E)$, as at the beginning of the game any state has appeared zero times, hence an even number of times. For a simple example let us consider
 that  $\{0,0,1,0,1,0,1,\cdots\}$ is the sequence of minority sides obtained in a game with $m=2$. Then if the standard updating rule is used, 
$\tilde{\mu}^{1}=00$, $\tilde{\mu}^{2}=01$, $\tilde{\mu}^{3}=10$, etc.,
and at time step $t=6$, the states $\mu_{1}=00$, $\mu_{2}=01$, $\mu_{3}=10$
and $\mu_{4}=11$ have appeared $1,3,2$ and $0$ times respectively,
hence ${\mathcal{P}}_{\mathcal{E}}^{7}=(O,O,E,E)$. We will drop $t$ when referring to a generic time step, using $\mathcal{P}_{\mathcal{E}}(\mu)$ instead. 
 
The PTD can be summarized in the following way: if at some time step, for some state $\mu$ we have
\textbf{\emph{$\mathcal{P}_{\mathcal{E}} (\mu)=O$}}, then in the  next (and hence even) appearance of $\mu$,
the outcome of the game is \emph{very likely} to be the opposite to that obtained
in the previous appearance of $\mu$. Broadly speaking, this dynamics is due to the fact that 
 on even appearances of $\mu$ crowds of agents will move
together to the side rewarded in the previous odd appearance.   When crowds emerge in the game,
their contribution to $z$ is very important. Furthermore, crowd
effects are the reason why $z$ is a large number in this region,
showing that fewer resources are allocated to the population as a whole. 

In a little more than a decade, there have been many attempts 
from different backgrounds to give a formal framework to the game, and to
analytically reproduce the results observed in computer simulations \cite{libro-MG}. 
Marsili et al.~proposed an analytical approach by resorting to sophisticated tools from
statistical physics, such as non-equilibrium stochastic dynamics \cite{marsili-cuenta-MG} and
the replica method \cite{MG-spin-glass}; Heimel et al.~proposed
an analytical approach based on the generating functional analysis \cite{heimel exact}. A geometric approach is based on the use of the {\it Reduced Strategy Space} defined as the subset of independent strategies plus the corresponding totally anticorrelated strategies \cite{zhang-1998,challet-zhang-scaling}. Based on this approach, Johnson et al.~derived approximate expressions
for the attendance fluctuations \cite{johnson, Hart-2001}.  In \cite{caridi-1} a mean field approach to the $MG$ is presented and the standard deviation of the key variable $N_{1}-N_{0}$
is computed by introducing a simplified framework. The
key idea in \cite{caridi-1} consists in defining a particular instance of the
game so that, for any value of $m$, all possible strategies and all
possible agents (each one represented by a possible pair of strategies in the case $s=2$) take part in the game. We call this particular instance the \emph{Full Strategy Minority Game} ($FSMG$).
For $s=2$, the number of {\it potential} 
agents is $\mathcal{N}={\mathcal{L} \choose 2}+\mathcal{L}$, 
where the first term represents all agents with two different strategies, and the  
second term represents the number of agents whose two strategies
are identical. Thus, the number of agents of the $FSMG$ is a function of $m$,
$\mathcal{N}=\mathcal{N}(m)$. Certain symmetries which appear only partially
in the $MG$ can be fully exploited in the $FSMG$, and this approach leads to interesting
theoretical results in the PTD region. 

In this paper, we show that the calculations given in \cite{caridi-1} can be highly simplified and   
easily applied to other variants of the $MG$, as the above-mentioned $MG_{rand}$
and $MG_{per}$. By taking advantage of its inherent symmetries, we analytically solve 
the $FSMG$. In particular, we  show the role of the number of states in odd 
occurrences in the calculation of the variable $z$ arriving to similar conclusions as that
given in \cite{tres-fases} by means of a different approach. 
 What is more, our calculations also apply to certain kinds of initial biased scores if the bias is introduced at the agents' level, i.e. given a positive $u_o$, any agent decides to assign $u_o$ virtual points "a priori" to any of their strategies with probability $p_b$. On the other hand
we define the Strict Period Two Dynamics (SPTD) as a
PTD with probability $1$, and show that SPTD is the characteristic dynamics of the $FSMG$. 
Moreover, by considering the $MG$ as a statistical sample of size $N$ of the $FSMG$, we are able to predict
the validity region of PTD for the $MG$ in a precise way. This theoretical result agrees fairly well with computer simulations.

\section{The Full Strategy Minority Game}

\subsection{Symmetry of the FSMG}
 
 Let ${\mathcal{H}}=2^{m}$, ${\mathcal{L}}=2^{\mathcal{H}}$ and ${\mathcal{N}}={\mathcal{L} \choose 2}+\mathcal{L}$
be the number of different states, different strategies and different
\textit{potential} agents for a $MG$ with parameter $m$. If we take
an arbitrary number $N$ of agents it is clear that, in randomly generated instances $I$ of the $MG$, the
following may happen: \emph{(1)} some potential agent may not participate
in the game or \emph{(2)} a multiple copy of the same agent may participate in the game.
In the Full Strategy Minority Game both cases are excluded: by construction 
the number of agents in the $FSMG$ is set to $\mathcal{N}$ and a single copy of every potential agent is allowed. 
  $\mathcal{S}_{\mathcal{H}}$ and $\mathcal{S}_{\mathcal{L}}$ (also known as the {\it Full Strategy Space})
  denote the set of states and strategies respectively. Throughout this article, the symbol $\sharp$
stands for the cardinality of a set, hence we have  $\sharp\mathcal{S}_{\mathcal{H}}={\mathcal{H}}$,
and $\sharp\mathcal{S}_{\mathcal{L}}={\mathcal{L}}$. For a given
state $\mu\in\mathcal{S}_{\mathcal{H}}$, the subset of strategies
in $\mathcal{S}_{\mathcal{L}}$ that predict a certain outcome $\tilde{o}$
for the state $\mu$ is denoted by $\mathcal{S}_{\mathcal{L},\mu\to \tos}$. For an arbitrary outcome $\tilde{o}\in\{0,1\}$ we will denote the opposite side by $\sim \tilde{o}$. It is clear that $\mathcal{S}_{\mathcal{L},\mu\to \tilde{o}}\cup\mathcal{S}_{\mathcal{L},\mu\to \sim \tilde{o}}=\mathcal{S}_{\mathcal{L}}$,
and that $\sharp\mathcal{S}_{\mathcal{L},\mu\to \tilde{o}}=\sharp\mathcal{S}_{\mathcal{L},\mu\to \sim \tilde{o}}=\mathcal{L}/2$. This means that for each state $\mu$, the number of strategies predicting $\tilde{o}$ and the number of strategies predicting $\sim \tilde{o}$ coincide. This symmetry together with the assumption of the SPTD  
allow us to make a remarkable analytic simplification of the game. Indeed, 
 let us consider  an arbitrary configuration $\mathcal{E}$ of the $FSMG$, and its state
sequence $\mathcal{M}$ (note that there exists only one instance $I$ for each $m$ in the $FSMG$). 
Suppose that at step $s$ a certain state $\tilde\mu^s=\mu$ 
 occurs for the first (and hence odd) time, and call $\tilde o$ the winning side after the voting round $s$. In that case  the parity array verifies 
 $ \mathcal{P}^{s+1}(\mu)=O$ (since $\mathcal{P}^l(\mu)=E$ if $l\le s$)  and strategies belonging to the set $\mathcal{S}_{\mathcal{L},\mu\to\tilde{o} }$ are rewarded
with a virtual point. Suppose now that at the time step $s'$,
$s'>s$, $\mu$ occurs for the second time 
(i.e., $\tilde{\mu}^{s'}=\mu$), then the SPTD implies that the winning side after the voting round $s'$ will be $\sim\tilde{o}$, and thus 
exactly the \textit{other half} of the strategies ($\mathcal{S}_{\mathcal{L},\mu\to \sim \tilde{o}}$) 
will correctly predict the minority side.
Therefore,  if we remove the point previously assigned to the set 
of strategies $\mathcal{S}_{\mathcal{L},\mu\to\tilde o}$   instead of adding a new point to
 strategies belonging to $\mathcal{S}_{\mathcal{L},\mu\to\sim \tilde o}$, the dynamics of the game will remain unchanged. Taking into account this rule, it is clear that the number of virtual points accumulated for any strategy, at any time step, ranges from $0$ to $\mathcal{H}$.  In equation
 (\ref{SPTD2}) we show that the $FSMG$ necessarily \emph{verifies} the SPTD, and hence the SPTD assumption can be removed in the previous argument. As far as we know this is the first example in the literature of a  game enjoying this property.

In the sequel we denote with $\mathcal{S}_{\mathcal{L},l}$
the set of strategies with $l$ virtual points, and with $\mathcal{S}_{\mathcal{L},\mu\to\tilde o,l}$
(resp.~$\mathcal{S}_{\mathcal{L},\mu\to \sim \tilde o,l}$) the set of strategies
with $l$ virtual points that predict $\tilde o$ (resp.~$\sim \tilde o$) for $\mu$.

\subsection{Strategies of the FSMG}

Given a parity array $\mathcal{P}_{\mathcal{E}}$ at an \textit{arbitrary}
time step,  we define   $n_e$ (resp.~$n_o$) as the number of even 
(resp.~odd)  symbols stored in $\mathcal{P}_{\mathcal{E}}$,   we have that
$0\le n_{e},n_{o}\le\mathcal{H}$ and $n_{o}=\mathcal{H}-n_{e}$. Below, we show that
the number $\sharp \mathcal{S}_{\mathcal{L},l}$ of strategies with \textit{exactly} $l$ virtual
points depends only on $\mathcal{P}_{\mathcal{E}}$, and for a given
state $\mu$ the values of $\sharp \mathcal{S}_{\mathcal{L},\mu\to\tilde{o},l}$ and $\sharp \mathcal{S}_{\mathcal{L},\mu\to\sim \tilde{o},l}$ depend only on $\mathcal{P}_{\mathcal{E}}$ and the parity of $\mu$.  

If $n_{e}=\mathcal{H}$,  
then no strategies can have virtual points due to the fact that \textit{only
after odd appearances of states} the strategies may get a point, i.e., $\sharp\mathcal{S}_{\mathcal{L},0}=\sharp\mathcal{S}_{\mathcal{L}}=\mathcal{L}$,
and $\sharp\mathcal{S}_{\mathcal{L},i}=0$ for $1\le i\le\mathcal{H}$.
On the other hand, if $n_{e}=\mathcal{H}-1$ then there is only one state,
say $\mu_{1}$, in an odd occurrence. In this case, strategies can have
virtual points due to this state only. But now, regardless of the outcome
$\tilde{o}$ (i.e., $\tilde{o}=0$ or $\tilde{o}=1$) after
this odd ocurrence of $\mu_{1}$, strategies belonging
to $\mathcal{S}_{\mathcal{L},\mu_{1}\to\tilde{o}}$ will have exactly
one virtual point, and those in the complement $S_{\mathcal{L},\mu_{1}\to\sim\tilde{o}}$
will have exactly $0$ points, hence $\sharp\mathcal{S}_{\mathcal{L},1}=\sharp\mathcal{S}_{\mathcal{L},0}=\mathcal{L}/2$,
and $\sharp\mathcal{S}_{\mathcal{L},i}=0$ for $2\le i\le\mathcal{H}$.
In the general case for parity numbers $n_{o}$ and $n_{e}$ we observe
that the \textit{maximum number of virtual points for a given strategy
is bounded by $n_{o}$}, since even ocurrences do not add any points.

Let $0\le l\le n_{o}$, and let $\mu_{k_{1}},\cdots,\mu_{k_{n_{o}}}$
be the states recording odd numbers of ocurrences, and $\tilde{o}_{1},\tilde{o}_{2},\cdots,\tilde{o}_{n_{o}}$
be the outcomes after the last ocurrence of each $\mu_{k_{i}}$. The number
$\sharp  \mathcal{S}_{\mathcal{L},l}$ can be easily computed by noting that a strategy
has exactly $l$ points if it has succesfully predicted the outcome
(i.e., it has won a virtual point) $l$ times out of the $n_{o}$ time steps corresponding to the last occurrences of the odd states,
and has \textit{unsuccesfully} predicted the outcome in the remaining $n_{o}-l$ time steps. For instance, if a strategy belongs to \[
\mathcal{F}_{\mu_{k_{1}},\cdots,\mu_{k_{l}}}=\mathcal{S}_{\mathcal{L},\mu_{k_{1}}\to\tilde{o}_{1}}\cap\mathcal{S}_{\mathcal{L},\mu_{k_{2}}\to\tilde{o}_{2}}\cdots\cap\mathcal{S}_{\mathcal{L},\mu_{k_{l}}\to\tilde{o}_{l}},\]
then it has succesfully predicted the outcome at least $l$ times. If in addition
it belongs to 
\[
\mathcal{F}_{\sim \mu_{k_{l+1}},\cdots,\sim \mu_{k_{n_{o}}}}=\mathcal{S}_{\mathcal{L},\mu_{k_{n_{l+1}}}\to\sim\tilde{o}_{n_{l+1}}}\cdots\cap\mathcal{S}_{\mathcal{L},\mu_{k_{n_{o}}}\to\sim\tilde{o}_{n_{o}}}\]
then it has unsuccesfully predicted the outcome at least $n_{o}-l$ times. So, the strategies
belonging to \[
\mathcal{F}_{\mu_{k_{1}},\cdots,\mu_{k_{l}},\sim \mu_{k_{l+1}},\cdots,\sim \mu_{k_{n_{o}}}}=\mathcal{F}_{\mu_{k_{1}},\cdots,\mu_{k_{l}}}\cap\mathcal{F}_{\sim \mu_{k_{l+1}},\cdots,\sim \mu_{k_{n_{o}}}}\]
 have exactly $l$ points. We easily find that 
 \begin{equation}
	 \sharp\mathcal{F}_{\mu_{k_{1}},\cdots,\mu_{k_{l}},\sim \mu_{k_{l+1}},\cdots,\sim \mu_{k_{n_{o}}}}=\mathcal{L}/2^{n_{o}} 
	 \label{paraines}
 \end{equation}
since only $\mathcal{L}/2$ strategies belong to $\mathcal{S}_{\mathcal{L},\mu_{k_{1}}\to\tilde{o}_{1}}$
and from those strategies only half succesfully predict the outcome
$\tilde{o}_{2}$ for the state $\mu_{k_{2}}$, i.e., $\sharp(\mathcal{S}_{\mathcal{L},\mu_{k_{1}}\to\tilde{o}_{1}}\cap\mathcal{S}_{\mathcal{L},\mu_{k_{2}}\to\tilde{o}_{2}})=\mathcal{L}/2^{2}$,
and so on until we obtain the number $\mathcal{L}/2^{l}$ of strategies
that have succesfully predicted $\tilde{o}_{1},\cdots,\tilde{o}_{l}$.
But from this set exactly half fails to predict $\tilde{o}_{l+1}$
(i.e., half of them predict $\sim\tilde{o}_{l+1}$) and so on. In this way we finally arrive to (\ref{paraines}). Since there are ${n_{o} \choose l}$
ways of choosing $l$ states from the set $\{\mu_{k_{1}},\cdots,\mu_{k_{n_{o}}}\}$
of states on some odd occurrence, we get
\begin{equation}
\sharp \mathcal{S}_{\mathcal{L},l}={n_{o} \choose l}\mathcal{L}/2^{n_{o}}.\label{lpuntos}
\end{equation}

 An important remark is the following: for a given state $\mu$ and
a given parity array $\mathcal{P}$, the set $\mathcal{S}_{\mathcal{L},l}$
can be decomposed in two sets as $\mathcal{S}_{\mathcal{L},l}=\mathcal{S}_{\mathcal{L},\mu\to\tilde{o},l}\cup \mathcal{S}_{\mathcal{L},\mu\to\sim\tilde{o},l}$. Whether or not  
these sets have the same cardinality \textit{depends on
the parity} of the state $\mu$. Indeed, if $\mu$ is on an even state  (i.e., $\mathcal {P}(\mu)=E$),  we have 
\begin{equation}
\sharp S_{\mathcal{L},\mu\to\tilde{o},l}=\sharp S_{\mathcal{L},\mu\to\sim\tilde{o},l}=\sharp S_{\mathcal{L},l}/2={n_{o} \choose l}\mathcal{L}/2^{n_{o}+1}\label{pares}\end{equation}
since we can apply the same argument used to get (\ref{lpuntos})
to the set $\mathcal{S}_{\LL,\mu\to\tilde{o}}$ (and $ \mathcal{S}_{\mathcal{L},\mu\to\sim\tilde{o}}$). Indeed, the set  $\mathcal{S}_{\LL,\mu\to\tilde{o}}$ has cardinality $\LL /2$ and the same symmetry properties than
 $\mathcal{S}_{\mathcal{L},l}$ (due to the fact that $\mathcal P(\mu)=E$). The situation is different if we take a state $\mu$ with $\mathcal{P}(\mu)=O$, in fact, the cardinality of the sets
$S_{\mathcal{L},\mu\to\tilde{o},l}$, $S_{\mathcal{L},\mu\to\sim\tilde{o},l}$ \textit{cannot} be equal. 
Indeed, if after the last  appearance of $\mu$ the
outcome of the game was $\tilde{o}$, then the strategies belonging to
$S_{\mathcal{L},\mu\to\tilde{o}}$ have \textit{at least} one virtual point,
so those strategies with $l$ points have successfully predicted the outcome exactly
$l-1$ times out of the $n_{o}-1$ time steps corresponding to the remaining states recording an
odd ocurrence, i.e., 
\begin{equation}
\sharp S_{\mathcal{L},\mu\to\tilde{o},l}={n_{o}-1 \choose l-1}\mathcal{L}/2^{n_{o}}\label{imparo}\end{equation}
and those belonging to $S_{\mathcal{L},\mu\to\sim\tilde{o},l}$ have successfuly
predicted the outcome exactly $l$ times out of the $n_{o}-1$ time steps corresponding
to all the states recording an odd ocurrence, i.e., \begin{equation}
\sharp S_{\mathcal{L},\mu\to\sim\tilde{o},l}={n_{o}-1 \choose l}\mathcal{L}/2^{n_{o}}.\label{imparno}\end{equation}
We again have $S_{\mathcal{L},\mu\to\tilde{o},l}+S_{\mathcal{L},\mu\to\sim\tilde{o},l}=S_{\mathcal{L},l}$,
which holds since ${n_{o}-1 \choose l-1}+{n_{o}-1 \choose l}={n_{o} \choose l}$.

\subsection{Agents and dynamics of the FSMG}
 
Let us focus on the $\mathcal N$ agents of the $FSMG$. With $\mathcal{N}_{\tilde o}$ and $\mathcal{N}_{\sim \tilde o}$ we denote the number of agents choosing the option  $\tilde o$ and the opposite option $\sim \tilde o$, respectively.  
When needed, we explicitly write $\mathcal N_{\tilde o}^t$ and $\mathcal N_{\sim \tilde o}^t$ to denote the dependence of these variables on the time step $t$.

For a given time step $t$ and a parity array $\mathcal{P}^t_{\mathcal{E}}$, we know from the previous section the distribution of virtual
points for strategies (\ref{lpuntos}). Recall that we have $n_e^t$ and $n_o^t$ states
on even and odd occurrences respectively for a given $\mathcal P^t_{\mathcal E}$ . We drop $t$ from $n_o$ and $n_e$ for the sake of
simplicity. Now we can predict  the polls result
if agents of the $FSMG$ are required to process a certain state $\mu_p$ (called the {\it present state}) given \textit{by an arbitrary updating rule}. To this end, we call {\it undecided
agents} to the agents that
have  {\it both strategies equally rewarded} and {\it predict opposite sides} for $\mu_p$,
i.e., one strategy belongs to $\mathcal S_{\mathcal L,\mu_p\to  \tilde o,l}$ and the other to $\mathcal S_{\mathcal L,\mu_p\to \sim \tilde o,l}$ , for certain $l$. 
The number of undecided agents is denoted by $\mathcal N_u$. On the other hand, {\it decided  agents} are the remaining ones, so that
$\mathcal N_d=\mathcal N- \mathcal N_u$. We also introduce the number of decided and undecided agents who choose the option $\tilde o$ to be $\mathcal N_{d_{\tilde o}}$ and $\mathcal N_{u_{\tilde o}}$, respectively. Clearly, we have
$$\mathcal N_{\tilde o}=\mathcal N_{d_{\tilde o}}+\mathcal N_{u_{\tilde o}} \qquad \mbox{and} \qquad \mathcal N_{\sim \tilde o}=\mathcal N_{d_{\sim \tilde o}}+\mathcal N_{u_{\sim \tilde o}}.$$

In the following, we will show that $\mathcal N_u$,  $\mathcal N_{d_{\tilde o}}$ and $\mathcal N_{d_{\sim \tilde o}}$ can be explicitly calculated for a given state $\mu_p$. 
The value $\mathcal N_u$ can be obtained by noticing that undecided agents have, for each $l$, one strategy from $\mathcal{S}_{\mathcal{L},\mu_p\to \tilde o,l}$
and the other one from $\mathcal{S}_{\mathcal{L},\mu_p\to \sim \tilde o,l}$. The total number of agents in this
situation is
\begin{equation}
 \label{indeciss}
 \mathcal N_u=\sum_{l=0}^{n_o}\sharp \mathcal S_{\mathcal L,\mu_p\to \tilde o,l} \sharp \mathcal S_{\mathcal L,\mu_p\to \sim \tilde o,l}.
\end{equation}
The sum ranges over all  possible virtual points that strategies may have for the given parity array  $\mathcal{P}_\mathcal{E}^t$.
On the other hand, decided agents that choose $\tilde o$ for the present state have
either two strategies from $\mathcal S_{\mathcal L,\mu_p\to \tilde o}$ (first and second terms in  equation (\ref{decisss})) or one strategy 
from  $\mathcal S_{\mathcal L,\mu_p\to \tilde o,l}$ and the other one 
from $\mathcal S_{\mathcal L,\mu_p\to \sim \tilde o,j}$, with $l>j$ (last term in (\ref{decisss})):
\begin{equation}
 \label{decisss}
 \mathcal N_{d_{\tilde o}}={ \sharp \mathcal S_{\mathcal L,\mu_p\to \tilde o}\choose 2 }+ \sharp \mathcal S_{\mathcal L,\mu_p\to \tilde o}+\sum_{l=1}^{n_o}
\sharp{\mathcal S_{\mathcal L,\mu_p\to \tilde o,l}}
\left( \sum_{j=0}^{l-1} \sharp {\mathcal S_{\mathcal L,\mu_p\to \sim \tilde o,j}}\right).
\end{equation}

 Let us first assume that $\mathcal P^t_{\mathcal E}(\mu_p)=E$, i.e., the present state $\mu_p$ has appeared an even number of times up to time step $t-1$. Suppose that now agents are required to process the state  $\mu_p$ (i.e., $\mu_p=\tilde \mu^t$), so this state is in an \emph{odd appearance}. Suppose now that $\tilde o$ was the outcome of the game after 
the last (and hence even) appearance of $\mu_p$. From (\ref{pares}), (\ref{decisss}) and 
(\ref{indeciss}) it is easy to check that $\mathcal N_{d_{\tilde o}}=\mathcal N_{d_{\sim \tilde o}}$, and that  

\begin{equation}
\mathcal N_u=\mathcal L^2/2^{2n_o+2}\sum_{l=0}^{n_o} {n_o\choose l}^2.
	\label{nupar}
\end{equation}
In short, if agents process a state in an \textit{odd} appearance, then we have
\begin{equation}
        \mathcal N_{d_{\tilde o}}-\mathcal N_{d_{\sim \tilde o}}=0, \qquad \mathcal N_u =\mathcal L^2/2^{2n_o+2}{2n_o\choose n_o}.
        \label{totalpar}
\end{equation}

Let us assume now that $\mathcal P^t_{\mathcal E}(\mu_p)=O$ (i.e., the present state is in an \emph{even appearance}). In this case, we use (\ref{imparo}) and (\ref{imparno}) to obtain $\mathcal{N}_u$, $\mathcal N_{d_{\tilde o}}$, and $\mathcal N_{d_{\sim \tilde o}}$:

\begin{equation}
\mathcal N_u=\mathcal L^2/2^{2n_o}\sum_{l=1}^{n_o-1} {n_o-1\choose l-1}{n_o-1\choose l}=\mathcal L^2/2^{2n_o}{2(n_o-1)\choose n_o},
	\label{nuimpar}
\end{equation}
that is obtained by using the Vandermonde's identity.

\begin{equation}
	\mathcal N_{d_{\tilde o}}= {\mathcal L/2 \choose 2} + \mathcal L/2 +
	\mathcal L^2/2^{2n_o}\left(\sum_{l=1}^{n_o}{n_o-1\choose l-1}\left(
	\sum_{j=0}^{l-1}{n_o-1\choose j} \right)  \right),
	\label{ndimparo}
\end{equation}

\begin{equation}
	\mathcal N_{d_{\sim \tilde o}}= {\mathcal L/2 \choose 2} + \mathcal L/2 +
	\mathcal L^2/2^{2n_o}\left(\sum_{l=1}^{n_o-1}{n_o-1\choose l}\left(\sum_{j=1}^{l-1}{n_o-1\choose j-1} \right)  \right).
	\label{ndimparno}
\end{equation}
In short, if agents process a state in an \emph{even appearance}, then 
\begin{eqnarray}
        \mathcal N_{d_{\tilde o}}-\mathcal N_{d_{\sim \tilde o}}&=&\mathcal L^2/2^{2n_o} {2n_o-1\choose n_o}, \nonumber \\ 
        \mathcal N_u&=&\mathcal L^2/2^{2n_o}{2(n_o-1)\choose n_o}.
                \label{totalimpar}
\end{eqnarray}
These equations lead to the following remarkable result
\begin{equation}
\mathcal N_{d_{\tilde o}}-\mathcal N_{d_{\sim \tilde o}}-\mathcal N_u = \mathcal L^2/2^{2n_o}{2(n_o-1)\choose n_o-1}>0,
	\label{SPTD}
\end{equation}
that says that  {\it FSMG verifies  necessarily} the SPTD when the present state is in an even occurrence. Indeed, (\ref{SPTD})   was obtained under the assumption that $\tilde o$
was the outcome of the game after the previous appearance of $\mu_p$ and shows that 
\begin{equation}
\mathcal N_{d_{\tilde o}}-\mathcal N_{d_{\sim \tilde o}}>\mathcal N_u
	\label{SPTD2}
\end{equation}
at the occurrence of the present state. Then, no matter what undecided agents do, the minority side will be $\sim \tilde o$ after the agents have processed the new appearance of $\mu_p$.

One may think that since SPTD was an assumption, then (\ref{SPTD2})  is not surprising. However, we can now remove such an assumption. In fact, let us consider
a configuration $\mathcal{E}$ of the $FSMG$, and its
sequence of states $\mathcal{M}=\{\tilde \mu_1,\tilde \mu_2,\cdots \}$. Since $\mathcal{P}^1_{\mathcal{E}}=(E,E,\cdots,E)$, in time step $1$  
agents process $\mu_1$, and we have  $\mathcal{P}^1_{\mathcal{E}}(\mu_1)=E$.  So we are on an {\it odd appearance} of $\mu_1$ and we easily see that our previous
arguments lead us to (\ref{totalpar}). The first repeated state will occur in at most $\HH$ steps  (the size of the state space), until then all 
states will occur in an  {\it odd} number of times and
{\it virtual points of strategies are bounded by $\HH$}, so we can still apply our arguments leading once more
to (\ref{totalpar}).  Since the first time that a state needs to be processed in an 
even occurrence will take place during the first $\HH+1$ steps, then the strategies' virtual points range from $0$ to $\HH$. Due to this fact, the previous arguments also apply for the first even appearance of a state, hence  (\ref{SPTD2}) holds and enforces the SPTD,  allowing us to apply the modified points assignment rule for the first time. By repeating the argument for the second, third, etc.~even appearances of states we can conclude the validity of the SPTD in general. 

\section{From the FSMG to the MG}
\subsection{Distribution of the agents' choices}
An instance $I$ of the $MG$ is a (random) sample of size $N$ from the $\mathcal N$ agents of the $FSMG$.
Consider an experiment consisting in randomly extracting (with reposition) a sample of size $N$ from a \emph{box} containing $\mathcal N$ agents of three different types, namely
$\mathcal N_{d_{\tilde o}}$ of type 1, $\mathcal N_{d_{\sim \tilde o}}$ of type 2 and $\mathcal N_u$ of type 3.
Suppose that after extracting the sample, we obtain
$N_{d_{\tilde o}}$ agents of type 1, $N_{d_{\sim \tilde o}}$ agents of type 2, and $N_u$ agents of type 3, so that

\begin{equation}
        N_{d_{\tilde o}}+N_{d_{\sim \tilde o}}+N_{u}=N.
\label{eq:total N}
\end{equation}
In this setting, $(N_{d_{\tilde o}}, N_{d_{\sim \tilde o}}, N_u)$ is a random variable with a multinomial probability distribution. Therefore, the probability of obtaining
$N_{d_{\tilde o}}$, $N_{d_{\sim \tilde o}}$ and $N_u$ is:

\begin{equation}
        P(x_1=N_{d_{\tilde o}},x_2=N_{d_{\sim \tilde o}},x_3=N_{u})=\frac{N!}{(N_{d_{\tilde o}})!\,(N_{d_{\sim \tilde o}})!\,(N_{u})!}
        p_1^{N_{d_{\tilde o}}}p_2^{N_{d_{\sim \tilde o}}}p_3^{N_{u}},
\label{multinomial}
\end{equation}
where
\begin{equation}
        p_1=\frac{\mathcal{N}_{d_{\tilde o}}}{\mathcal{N}}, \ \ p_2=\frac{\mathcal{N}_{d_{\sim \tilde o}}}{\mathcal{N}}, \ \  p_3=\frac{\mathcal{N}_{u}}{\mathcal{N}}.
        \label{laspes}
\end{equation}

Our aim is to characterize the variable $N_{\tilde o}$ (i.e., the number of agents who choose  side $\tilde o$). To this end, consider a second experiment: once a sample with
$N_{d_{\tilde o}}$, $N_{d_{\sim \tilde o}}$ and $N_u$ agents of each type is obtained, we need to know the distribution of $N_{u_{\tilde o}}$, i.e., the number of undecided agents choosing $\tilde o$. Since undecided agents  randomly
pick among  their strategies, a reasonable assumption is that $N_{u_{\tilde o}}\sim Bi(N_u,1/2)$. From $N_{\tilde o}=N_{d_{\tilde o}}+N_{u_{\tilde o}}$ one can easily  show that 
$$  N_{\tilde o}\sim Bi(N,p_1+p_3/2)$$
or, using that $p_1+p_2+p_3=1$,
\begin{equation}
 N_{\tilde o}\sim Bi(N,1/2+(p_1-p_2)/2).
 \label{distrino}
 \end{equation}
 Equation (\ref{distrino}) will be useful in the following sections.

\subsection{Analytic calculations for different updating rules}
\label{secanalitica}
For a fixed side $\tos$, the expected value 
\begin{equation}
	\sigma^2 =\ \ <(N_{\tilde o}- N/2)^2>
	\label{esperanzass}
\end{equation}
over every possible configuration $\mathcal{E}$  measures the
waste of the population's resources. Our previous calculations
allow us to compute the expected values on  odd $\sigma^2_o=\ <(N_{\tilde o}- N/2)^2>_o$ and even $\sigma^2_e =\ <(N_{\tilde o}- N/2)^2>_e$  occurrences of the present state $\mu_p$, in order to find a closed expression  for (\ref{esperanzass}).

Given a random variable $y\sim 
Bi(N,p)$ it is known that $Var(y)=Np(1-p)$, and $<y>=Np$,  hence from (\ref{distrino}) we have
that 
\begin{eqnarray}
\label{simplevar}
Var(N_{\tos})=N/4(1-(p_1-p_2)^2)=N/4\left(1-\left(\frac{\mathcal{N}_{d_{\tos}}-\mathcal{N}_{d_{\sim \tos}}}{\mathcal{N}}\right)^2\right), \nonumber  \\
<N_{\tos}>=N/2(1+(p_1-p_2))=N/2\left(1+\left(\frac{\mathcal{N}_{d_{\tos}}-\mathcal{N}_{d_{\sim \tos}}}{\mathcal{N}}\right)\right).
\end{eqnarray}
where the last identity  follows from  (\ref{laspes}). By (\ref{totalpar}) and (\ref{totalimpar}), we have  that $ \mathcal{N}_{d_{\tos}}-\mathcal{N}_{d_{\sim \tos}}$ only depends on $n_o$ for both odd and
even appearances of $\mu_p$. Therefore, if for a given {\it updating rule}  the probability
distribution of the random variable $n_o$ is known, then  (\ref{esperanzass})
can be computed. 

Clearly, we have
\begin{equation}
	\sigma^2 =Var(N_{\tos}-N/2)+(<N_{\tilde o}>- N/2)^2=Var(N_{\tos})+(<N_{\tilde o}>- N/2)^2.
	\label{esperanzassen2}
\end{equation}
By replacing (\ref{simplevar}) in (\ref{esperanzassen2}) we get
\begin{equation}
	\sigma^2 =N/4\left(1+(N-1)\left(\frac{\mathcal{N}_{d_{\tos}}-\mathcal{N}_{d_{\sim \tos}}}{\mathcal{N}}\right)^2\right).
	\label{esperanzassen3}
\end{equation}
We can use this equation on odd ($\sigma^2_o$) and even ($\sigma^2_e$)  occurrences. In odd occurrences   of the present state $\mu_p$
equation (\ref{totalpar}) yields $\mathcal{N}_{d_{\tos}}-\mathcal{N}_{d_{\sim \tos}}=0$ and hence   (\ref{esperanzassen3}) gives
\begin{equation}
 \label{impsiempre}
 \sigma^2_o=N/4
\end{equation}
{\it regardless the updating rule}, as observed when the variable $\sigma^2_o/N$ is computed in numerical simulations. On {\it even} occurrences of $\mu_p$ --when the crowd effect arises-- from (\ref{esperanzassen3}) and (\ref{totalimpar}) we obtain 
$$ \sigma^2_{e,n_o}=N/4\left(1+(N-1)\left( \frac{\mathcal L^2/2^{2n_o} {2n_o-1\choose n_o}}{\mathcal{N}} \right)^2\right)$$
for a parity array with $n_o$ states in odd ocurrences. Using that $\frac{\mathcal L^2}{\mathcal N}=\frac{2}{1+\mathcal{ L}^{-1}}$, $\mathcal L=2^{\mathcal H}$ and ${2n_o-1\choose n_o}={2n_o\choose n_o}/2$, we can write

\begin{equation}
 \label{parsiempreno}
 \sigma^2_{e,n_o}=N/4\left(1+\frac{N-1}{(1+2^{-\mathcal H})^2} \left( 1/2^{2n_o} {2n_o\choose n_o} \right)^2\right).
\end{equation}
The role of the updating rule becomes relevant in the ``typical'' numbers $n_o$ that appear when agents are processing a state $\mu_p$ in an {\it even} occurrence. To be more precise, if $p(n_o|\mu_p\equiv e)$ stands for the probability of finding $n_o$ states in odd occurrence given that $\mu_p$ is on an even 
occurrence then we have  
\begin{equation}
 \label{parsiempre}
 \sigma^2_{e}=\sum_{n_o} p(n_o|\mu_p\equiv e) \sigma^2_{e,n_o}.
\end{equation}
We apply the obtained results to different kinds of updating rules. Our two
first examples are based on the exogenous updating rules found in $MG_{rand}$ \cite{cavagna-irrelevance-memory}  and $MG_{per}$ \cite{MG-per}. In the end of the subsection we address  the standard $MG$. Since the $FSMG$ verifies the SPTD and our calculations rely on the $FSMG$, {\it we expect to find good results only in the region of validity of the PTD}.

\begin{figure}[h]
\centering\includegraphics[width=6cm]{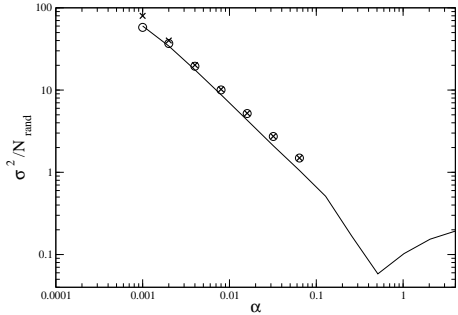}
\caption{The straight line shows $\sigma^2/N$ as a function of $\alpha$ for the $MG_{rand}$ for different values of $m$ (from 2 to 14) and $N=4001$. For each value of $N$ and $m$, 100 runs have been performed, each one of $T=100000$ time steps discarding the first 50000 steps. Empty circles show the analytical result (\ref{totalmgrand}), and star-shaped symbols show the approximated invariant (i.e., depending only on $\alpha$) expression (\ref{invariantemgrand}).}\label{fig2}
\end{figure}

In the $MG_{rand}$, the present state $\mu_p$  is  chosen at random (uniformly) from the whole set of states $\mathcal{S}_{\mathcal H}$. In this case one can easily get
\begin{equation}
 \label{condicrand}
p(n_o|\mu_p\equiv e)_{MG_{rand}}=1/2^{\mathcal H -1}{\mathcal H -1 \choose n_o -1}. 
\end{equation}
 Indeed, if $\mu_p$ is in an {\it even} appearance, then its parity is recorded as odd in $\mathcal P$, i.e.,
$\mathcal P(\mu_p)=O$, hence from the $\mathcal H-1$ remaining states $\mu$ we need to find the probability that exactly $n_o-1$ of them verify $\mathcal P(\mu)=O$. Since  each state is in odd occurrence with probability $1/2$, (\ref{condicrand}) follows. Now, we are ready to find an approximate value of $\sigma^2_{MG_{rand}}=1/2\sigma_o^2+1/2\sigma_e^2$. From (\ref{impsiempre}), (\ref{parsiempreno}), (\ref{parsiempre}), and (\ref{condicrand}), we obtain
\begin{equation}
\label{totalmgrand}
 \sigma^2_{MG_{rand}}/N=1/4+\frac{N-1}{8(1+2^{-\mathcal H})^22^{\mathcal H -1}}\sum_{n_0=1}^{\mathcal H} \left( 1/2^{2n_o} {2n_o\choose n_o} \right)^2{\mathcal H -1 \choose n_o -1}.
\end{equation}
It is not clear whether or not (\ref{totalmgrand}) depends only on $\alpha=2^m/N=\mathcal H/N$. However,
it can be approximated by (\ref{invariantemgrand}), which only depends on $\alpha$ (i.e., it follows the scaling
relation $\sigma^2_{MG_{rand}}/N \sim f(\alpha)$, \cite{chino-simu-largas}). Indeed,  due to the fact that in the $MG_{rand}$ approximately half of states will be in odd
occurrences, we can only take one ``typical'' value $n_o = \mathcal H/2$ in (\ref{parsiempreno}),
neglecting the contribution of any other $\sigma_{e,n_o}^2$ to $\sigma^2_e$ in   (\ref{parsiempre}). By doing this we obtain the following approximation:
$$\sigma^2_{MG_{rand}}/N \sim 1/4+ \left(\frac{1}{2^{\HH}(1+\frac{1}{2^{\HH}})}{\HH \choose \HH/2} \right)^2(N-1)/8. $$
Taking into account that $N\gg 1$, $2^{\HH}=2^{2^{m}}\gg 1$ (so that $N-1\sim N$, $1+\frac{1}{2^{\HH}}\sim 1$) , and using the well known approximation (which can be obtained straightforwardly from Stirling's formula),
$\frac{1}{2^{\HH}}{\HH \choose \HH/2}\sim \sqrt{\frac{2}{\HH \pi}}$ we get
\begin{equation}
	\sigma_{MG_{rand}}^2/N\sim 1/4+ \frac{N}{4 \HH \pi}.
	\label{invariantemgrand}
\end{equation}
In Figure \ref{fig2} we show the analytical result given by (\ref{totalmgrand}), the approximated expression (\ref{invariantemgrand}), and the numerical results for the $MG_{rand}$ with $N=4001$. 
\bigskip

\begin{figure}[h]
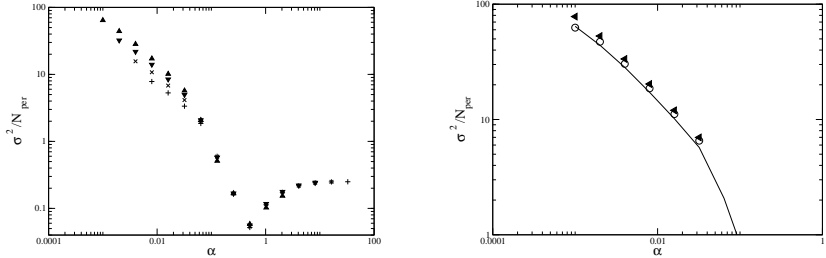

\centering\includegraphics[width=5cm]{fig3_v2.eps} \qquad \includegraphics[width=5cm]{fig3f_v2.eps}
\caption[loftitle]{On the left, $\sigma^2/N$  as a function of $\alpha$ for the $MG_{per}$ for different values of $m$ (in the range from 2 to 14) and $N=$: $+$ symbols for $N=501$; $\times$ $N=1001$; $\blacktriangledown $ $N=2001$; $\blacktriangle $ $N=4001$. For each value of $N$ and $m$ we perform 100 runs, each one of $T=100000$ time steps discarding the first 50000 steps. On the right, the full line is the $MG_{per}$ with $N=4001$, empty circles are given by (\ref{sigmatotalmgper}), and
$\blacktriangleleft $ show the approximated expression (\ref{igualchin}).   } \label{fig3}
\end{figure}

 Our next aim is to apply our calculations to the $MG_{per}$ introduced in \cite{MG-per}. In this
 case the updating rule follows a periodic pattern of period $\HH$ that runs over all the states.
 In fact, with our convention that identifies any state $\mu$ with its binary expansion plus one,  
 the updating rule proposed in \cite{MG-per} follows the natural order $1\to 2 \to 3 \cdots$ modulo
 $\HH$. In this case we have again $\sigma^2_{MG_{per}}=1/2\sigma_o^2+1/2\sigma_e^2$. On the other hand we clearly get 
 
\begin{equation}
 \label{condicper}
 p(n_o|\mu_p\equiv e)_{MG_{per}}=1/\HH
\end{equation}
 and then, from (\ref{condicper}) together with  (\ref{impsiempre}), (\ref{parsiempreno}), and (\ref{parsiempre}) we find
 
\begin{equation}
\label{sigmatotalmgper}
 \sigma_{MG_{per}}^2/N=1/4+ \frac{(N-1)}{8\HH (1+2^{-\HH})^2} \sum_{n_o=1 }^{\HH}
	\left(\frac{1}{2^{2n_o}}{2n_o \choose n_o}\right)^2.
\end{equation}
In Figure \ref{fig3} we show the agreement between (\ref{sigmatotalmgper}) and numerical experiments
for the $MG_{per}$ with $N=4001$. In much the same way as for the $MG_{rand}$, this expression can be highly simplified. Indeed, taking into account that $N\gg 1$, $2^{\HH}=2^{2^{m}}\gg 1$, and using again that 
 $\frac{1}{2^{2n_o}}{2n_o \choose n_o}\sim \sqrt{\frac{1}{n_o \pi}}$ , we get  
	$$\sigma_{MG_{per}}^2/N\sim 1/4+ \frac{N \sum_{n_o=1 }^{\HH}
	 1/n_o}{8\HH \pi}$$
that can be simplified once more
\begin{equation}
 \label{igualchin}
 \sigma_{MG_{per}}^2/N\sim 1/4+ \frac{N}{8\HH \pi}(\log \HH+\gamma),
\end{equation}
by using that  $\lim_{M\to \infty} \sum_{1\le i \le M} 1/i -\log (M)\to \gamma$, where  $\gamma=0.57\dots$ is the constant of  Euler-Mascheroni. Equation (\ref{igualchin}) is essentially the same obtained in
\cite{tres-fases} by means of a different  approach.

\begin{figure}[h]
\centering\includegraphics[width=4cm]{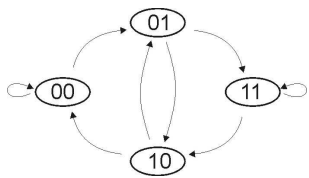}
\caption{De Bruijn diagram of order 2. All 4 possible $\mu$ states are represented: $00$, $01$, $10$ and $11$. Links represent the possible sequences among these states under the rules of the $MG$. For example, after the state $00$ there are two possible states which can follow: $01$ and $00$, and there are two possible preceding states: $10$ and $00$.} \label{fig5}
\end{figure}

Finally, we turn our attention to the standard $MG$. As before, the key tool is equation (\ref{parsiempre}), for which we need to know $p(n_o|\mu_p\equiv e)_{MG}$. We obtain this probability as follows. The 
updating rule in the standard $MG$ consists in defining the present state as the last $m$ outcomes of the game,
in this way any chain of correlative states given by the updating rule can be seen as a walk on the De Bruijn diagram corresponding to each $m$ \cite{challet-relevance-memory} (Figure \ref{fig5} shows the diagram of order 2). 

\begin{figure}[h]
\centering\includegraphics[width=8cm]{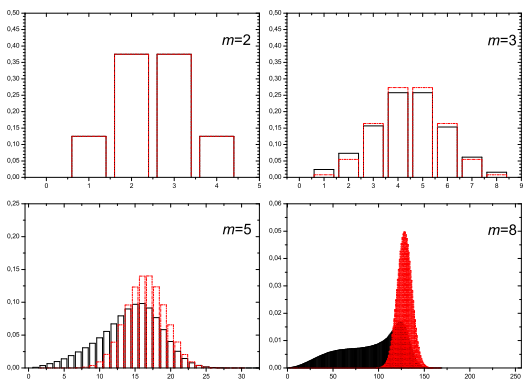}
\caption{Histograms of probability distribution $p(n_o|\mu_p\equiv e)_{MG}$  (line) and $p(n_o|\mu_p\equiv e)_{MG_{rand}}$  (dash) with parameters $m=2$, $m=3$, $m=5$ and $m=8$. These results were obtained by simulating the possible sequence of states under two assumptions: the sequence of states satisfies the rule of the $MG$ (i.e., is a walk in the De Bruijn diagram), and SPTD is valid in the even occurrence of each state. The $MG_{rand}$ is displayed for illustrative purposes, since it is analytically given by (\ref{condicrand}). } 
\label{fig6}
\end{figure}

Note that each node in the De Bruijn diagram has two incoming links and two outgoing links. We performed numerical random walks obeying the SPTD on the De Bruijn diagram, i.e., in odd occurrences of each state we choose at random one from the two available states allowed by the diagram, and on even occurrences we choose the state opposite to the one selected in the previous odd appearance. 
We recorded at each time the value of $n_o$ and after a simulation of 50 different walks of 500000 steps, we obtained 
the approximated probability distribution of the variable $n_o$, that can  be seen in Figure  \ref{fig6}.

\begin{figure}[h]
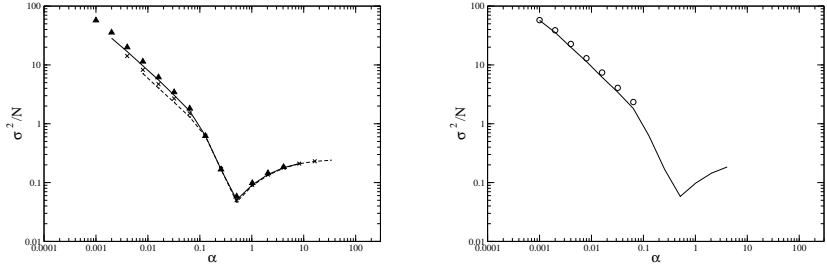

 \centering\includegraphics[width=5cm]{fig4a_v2.eps} \qquad \includegraphics[width=5cm]{fig4b_v2.eps}
\caption{Left: $\sigma^2/N$  as a function of $\alpha$ in the $MG$ for different values of $m$ (in the range from 2 to 14) and $N$. The dash line shows the case $N=501$; $\times$ $N=1001$; the full line shows $N=2001$; $\blacktriangle $ $N=4001$. For each value of $N$ and $m$ we performed 100 runs, each one of $T=100000$ time steps discarding the first  50000 steps. Right: 
Numerical MG (the full line), and analytical result ($\circ$) for $N=4001$.} \label{fig4}

\end{figure}

Using the obtained $p(n_o|\mu_p\equiv e)_{MG}$, we computed $\sigma^2_{MG}$ by means of 
 (\ref{impsiempre}), (\ref{parsiempreno}) and (\ref{parsiempre}). Figure \ref{fig4} shows the analytical and numerical results. 

\begin{figure}[h]
\centering\includegraphics[width=7cm]{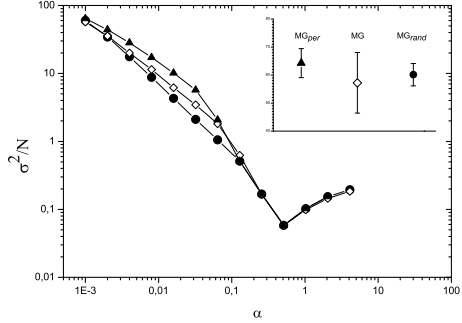}\caption{  The main figure shows $\sigma_{rand}^2/N=x_1$ (circles),  $\sigma^2_{per}/N=x_2$ (triangles) and $\sigma_{MG}^2/N=x_3$ (diamonds) for games with $N=4001$ and different values of $m$ in the range from 2 to 14. We simulated 100 independent runs of 100000 time steps each. We average the mean value of $\sigma^2/N$ using the last 50000 steps for each run. The figure inset reports the same values corresponding to $m=2$ with error bars ($\Delta x_{1}$, $\Delta x_{2}$ and $\Delta x_{3}$, respectively). The bars overlap for this value of $m$. We performed a $\chi^2$ statistical test on means in order to check the null hypothesis H0: ``the three values are mutually consistent within these error bars, i.e., the obtained values of $x_1$, $x_2$ and $x_3$ correspond to measurements of variables with the same mean value, and normal distribution'' \cite{frodesen}. We computed the statistical value $S$ as the weighted sum of the squared deviations from the weighted average value of the three measurements ($\overline{x}$), i.e., $S=\sum_{i=1}^{3} {(x_i-\overline{x})^{2}/\Delta x_{i}^2}$ where $x_i$ and $\Delta x_{i}$ are the variable and error corresponding to the $i$-th model and $\overline{x}$ is the maximum likelihood estimator of the mean of the three values. For $m=2$, we obtained $S=1.30$ and for this reason  H0 cannot be rejected, even when $p(n_o|\mu_p\equiv e)_{MG_{per}}$ clearly differs from $p(n_o|\mu_p\equiv e)_{MG}$ and  $p(n_o|\mu_p\equiv e)_{MG_{rand}}$. If we perform the test only for the couple $MG$ and $MG_{rand}$, we cannot reject H0 (in this case H0: ``the two values are mutually consistent'') for $m=3$, but we can reject it for $m=4,\dots,9$ with significance level smaller than $0.5 \%$.  For $m=3$ and the three models we obtain $S=20.54$ and, in this case, we reject H0 with less than $0.5 \%$ of significance level, concluding that results from $MG_{per}$ are different from $MG$ and $MG_{rand}$ in games with $m=3$ with a $0.5 \%$ probability of error of the second kind. 
}
\label{fig7}
\end{figure}

The fact that the histograms given in Figure \ref{fig6} agree for $m=2$ is particularly interesting. This  is
consistent with numerical simulations in which  $\sigma^2_{MG}\sim  \sigma^2_{MG_{rand}}$ for $m=2$ and $N=4001$ (see Figure \ref{fig7}). For $m=3$, the histograms corresponding to $MG$ and $MG_{rand}$ slightly differ but still are very similar, and a $\chi^2$ statistical test cannot reject the null hypothesis H0: ``the values obtained from $MG$ and $MG_{rand}$ are mutually consistent (i.e., correspond to variables with the same mean value)''.

 \medskip

\medskip

\subsection{Biased Initial Scores}
\label{bisc}

Our previous calculations were developed on the assumption that the strategies have zero virtual
points at the beginning of the game. As another application of the  $FSMG$, we consider the case in which 
an initial bias is introduced to the strategies scores. The first possibility could be the following
(see Ref \cite{galla}):
take a given $u_o>0$ and randomly assign $u_o$ (instead of zero) virtual points to
certain  strategies belonging to the whole set of strategies of the $FSMG$.
Following this idea, one is tempted to compute a variant of (\ref{lpuntos}) taking into account
such a bias. Regretfully, one can readily discover that the original symmetry of the $FSMG$ has been broken,
and it is not clear how to obtain an analogous of (\ref{lpuntos}). However, a second
possibility can be managed. Suppose that the bias is introduced {\it at the agents' level}, i.e. 
any agent randomly chooses, with a bias probability $p_b$, to assign
"a priori" $u_o$ virtual points to each one of their strategies. This way of assigning
initial scores differs from that studied in \cite{libro-MG,heimel exact,marsili-cuenta-MG}, where a uniform initial bias (called $y_0$
or $q(0)$ in those works) 
is applied to each agent. It is important to notice that,  in our approach at most a fraction $2p_b(1-p_b)$ (in average) of agents
have different initial scores (in fact a difference of $u_o$ points) between their strategies.

\begin{figure}[h]
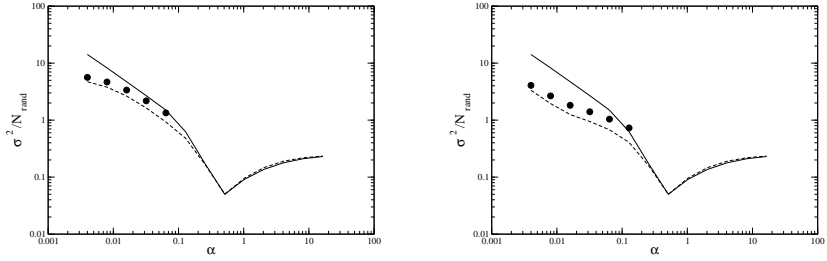

 \centering\includegraphics[width=5cm]{fig8a.eps} \qquad \includegraphics[width=5cm]{fig8b.eps}
\caption{Left: $\sigma^2/N$ as a function of $\alpha$. Numerical simulations of the unbiased $MG_{rand}$ (full line), and biased $MG_{RAND}$ with $u_o=2$ (dashed line).  Our analytical calculation is displayed in full circles.
Right: the same as left with $u_o=4$. In all the cases, $N=1001$, and for each value of $N$ and $m$ we performed 100 runs, each one of $T=100000$ time steps discarding the first  $50000$ steps. } \label{figuextra}

\end{figure}

For the sake of simplicity, we  fix the bias probability $p_b=1/2$  although we explain below how to handle more general values (in fact, rational 
values of $p_b$). Let us introduce an "enlarged" set of strategies  consisting in two copies
$\mathcal{S}_{\mathcal{L}}^1, \mathcal{S}_{\mathcal{L}}^2$ of
the original set of strategies $\mathcal{S}_{\mathcal{L}}$ of the $FSMG$. In 
$\mathcal{S}_{\mathcal{L}}^1$, all the strategies have $0$ virtual points and $u_o$ 
in $\mathcal{S}_{\mathcal{L}}^2$. Now, the {\it extended} $FSMG^E$ consists in a single copy
of any possible agent obtained from the enlarged set $\mathcal{S}_{\mathcal{L}}^{1,2}=\mathcal{S}_{\mathcal{L}}^1 \cup \mathcal{S}_{\mathcal{L}}^2$. Following our previous ideas, the described biased $MG$ can be thought as a statistical sample of the $FSMG^E$. The number of agents
of the $FSMG^E$ is now ${\mathcal{N}^E}={2\mathcal{L} \choose 2}+2\mathcal{L}$. The important fact in this scenario is that the symmetry still holds inside each $\mathcal{S}_{\mathcal{L}}^i$, and thanks to this one can replicate our previous
arguments obtaining the analogous of (\ref{lpuntos})
\begin{equation}
\sharp \mathcal{S}_{\mathcal{L},l}^{1,2}=
\sharp \mathcal{S}_{\mathcal{L},l}^{1}+\sharp \mathcal{S}_{\mathcal{L},l}^{2} ={n_{o} \choose l}\mathcal{L}/2^{n_{o}}+{n_{o} \choose l-u_o}\mathcal{L}/2^{n_{o}} .\label{lpuntosgen}
\end{equation}
 where $l$ ranges now from $0$ to $\HH+u_o$ (notice that the combinatorial number ${a\choose b}$
 is zero if $b>a$ or $b<0$). Therefore, analogous for (\ref{imparo}) and (\ref{imparno}) can be computed
 and hence the same thing can be done with (\ref{decisss}). By doing this, one can get for the enlarged 
 sets,
\begin{equation}
	 \mathcal N_{d_{\tilde o}}^E-\mathcal N_{d_{\sim \tilde o}}^E=\mathcal L^2/2^{2n_o} \left(  
 2{2n_o-1\choose n_o-1}+{2n_o-1\choose n_o-1+u_o}+{2n_o-1\choose n_o-1-u_o}\right)
 \label{laultima}
\end{equation}
$$
        \mathcal N_u^E=\mathcal L^2/2^{2n_o}   
        \left(2 {2(n_o-1)\choose n_o} + {2(n_o-1) \choose n_o-u_o} +  {2(n_o-1)\choose n_o+u_o}  \right) .
$$
when agents are processing  a state in an \emph{even appearance}. Furthermore, 
from these expressions, one can easily check that  the $SPTD$  is still verified in this new game.  Therefore, proceeding as before we can use   
(\ref{esperanzassen3}) in order to get analytical approximations for $\sigma^2/N$. In Figure \ref{figuextra} we compare our analytical calculations with numerical simulations of the $MG_{rand}$
in which any agent chooses to assign, with probability $1/2$, $u_o$ virtual points (instead of zero) with $u_o=2$ and $u_o= 4$ to their strategies. The calculations were carried out by using  (\ref{condicrand}), as the conditional
probabilities $p(n_o|\mu_p\equiv e)$. Comparing $ \frac{\mathcal N_{d_{\tilde o}}^E-\mathcal N_{d_{\sim \tilde o}}^E}{\mathcal N^E}$ with the unbiased $ \frac{\mathcal N_{d_{\tilde o}}-\mathcal N_{d_{\sim \tilde o}}}{\mathcal N}$, it is easy to prove that the bias implies a mitigation of the crowd effect, a fact that
becomes apparent in Figure \ref{figuextra}.
 \medskip

\medskip

In order to find expressions for similar games with $p=a/b$ (a rational number with $a,b$ integer
and positive numbers with $b\ge a$) then one needs to use an extended set of strategies in the following way:
take $b-a$ copies of the set   $\mathcal{S}_{\mathcal{L}}$ and keep the strategies in zero virtual
points, then take $a$ copies of the set   $\mathcal{S}_{\mathcal{L}}$ and  assign $u_o$ virtual points
to their strategies. Consider the enlarged $FSMG^E$ by taking a single copy of any possible agent
with strategies taken from copies of $\mathcal{S}_{\mathcal{L}}$. Now, the $p_b$-biased $MG$ can be understood as a sample of this game, and the calculations can be repeated step by step. Moreover,
different $u_o$ can be introduced in the same game by adding more copies of $\mathcal{S}_{\mathcal{L}}$.

Since the influence of initial biased scores of the type described in this section does not seem to be too well documented in the existing literature, we illustrate the behaviour of the biased game 
by means of several numerical simulations given in Section \ref{numres}.

\subsection{Range of validity of PTD}
\begin{figure}[h]
\centering\includegraphics[width=5cm]{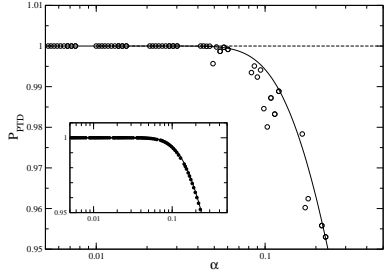}
\caption{Probability $P_{PTD}$ for PTD to take place in even occurrences of the states  as a function of $\alpha$, calculated as one minus the probability of breaking the PTD for \emph{the first time in a given simulation of the game}.  The full line is the analytic case (by using the approximation $n_o=\mathcal H/2$). Empty circles: results of the simulation of the $MG$. In the simulations we have computed $P_{PTD}$ as follows: at each even occurrence of each state, if after the poll the PTD is not fulfilled   (i.e., the minority side agrees with that obtained after the previous --odd-- occurrence of the same state), then we consider that this step does not contribute to $P_{PTD}$. At the same time, present state and virtual points are assigned as if the PTD had not failed. This way we compute the probability of breaking the PTD for \emph{the first time in the game}. The empty circles show the average of 100 runs of 100000 steps each, for $N=$ 501, 531, 561, 591, 621, 651, 681, 711, 741, 771 and $m=2,\dots,7$. Inset: comparison between 
the expression (\ref{aproxptdfin}) (the full line) and the analogous obtained by using the complete distribution
of $p(n_o|\mu_p\equiv e)$ given by (\ref{condicrand}) (full circles). Both curves are indistinguishable. }
\label{fig1}
\end{figure}

If $\tos$ was the outcome in the last odd occurrence of some state $\mu$, then for the next (hence even) appearance of $\mu$, equation (\ref{SPTD2}) shows that the $FSMG$ verifies the SPTD.  Since the $MG$ can be regarded as a statistical sample  of the $FSMG$, this brings a natural way to approximate  the probability of breaking the PTD for the first time in the $MG$. Indeed, if we denote with $P_{PTD}$  the probability of
verifying the PTD for some configuration $\mathcal E$ of the $MG$ we clearly have
$$P_{PTD}=P(N_{\tos}>N/2).$$
On the other hand, from (\ref{distrino}) we see that $P_{PTD}$ can be easily calculated. In fact, by using
the normal distribution approximation to the binomial (\ref{distrino}) we get
\begin{equation}
P_{PTD} \sim 1-\Phi\left(-\frac{\sqrt{N}(p_1-p_2) }{\sqrt{1-(p_1-p_2)^2}}\right),
 \label{aproxptd}
\end{equation}
where $\Phi$ stands for the cumulative function of the standard normal distribution. Probabilities $p_i$ can now be  calculated by means of (\ref{laspes}) and using  (\ref{totalimpar}). Since
(\ref{totalimpar}) depends on $n_o$, the values of $p(n_o|\mu_p\equiv e)$ are needed in order to
find $p_1-p_2$. For the $MG$ these values were obtained in Subsection \ref{secanalitica} by means of
the de Bruijin graph, but since we do not have an analytic expression for them, we may use the values  (\ref{condicrand}) for the $MG_{rand}$  {\it as an approximation}. Moreover, a further approximation
can be made, as it was mentioned before in the $MG_{rand}$ approximately half of states will be in odd
occurrences, and therefore we can take $n_o = \mathcal H/2$ as a  ``typical'' value (in fact the inset of Figure \ref{fig1} shows the analytical results for both approaches and there are not any noticeable differences).

Taking a rough approximation by using only the typical $n_o=\HH/2$ we arrive to the compact expression
 $$  
 P_{PTD}\sim 1-\Phi\left(\frac{ -\sqrt{N} \frac{1}{2^{\HH}(1+\frac{1}{2^{\HH}})}{\HH \choose \HH/2}}{
 \sqrt{1-\left(\frac{1}{2^{\HH}(1+\frac{1}{2^{\HH}})}{\HH \choose \HH/2} \right)^2}}\right)
 $$
and using that $1+\frac{1}{2^{\HH}}\sim 1$ and $\frac{1}{2^{\HH}}{\HH \choose \HH/2} \sim \sqrt{
\frac{2}{\HH \pi}}$ we get
\begin{equation}
P_{PTD} \sim 1-\Phi\left(-\sqrt{\frac{N}{ \frac{\HH\pi}{2}-1} } \right).
 \label{aproxptdfin}
\end{equation}
It is important to note that our arguments are developed under the assumption of SPTD. Once the period two dynamics is broken, the system falls into states whose values of $p_1$ and $p_2$ can not be calculated by means of this approach.
This means that Eq. (\ref{aproxptdfin}) is a good approximation for the probability of breaking PTD {\it for the first time} in a given realization  of the $MG$.
Despite the involved approximations, and as it is shown in Figure \ref{fig1}, theoretical results are in very good agreement with numerically computed values of $P_{PTD}$ for the $MG$.
\medskip

\section{Numerical simulations for the biased $MG$}
\label{numres}
\begin{figure}[h]
\centering\includegraphics[width=5cm]{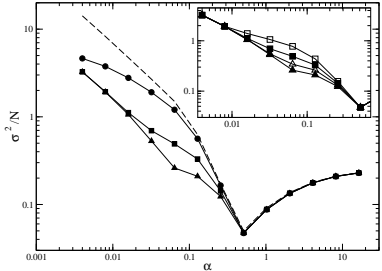}
\caption{$\sigma^2/N$ as a function of $\alpha$. Numerical simulations of the unbiased $MG$ (dash line), and biased $MG$ with $u_o=2$ (filled circles), $u_o=6$ (filled squares), and $u_o=10$ (filled triangle). The inset shows results for values of $u_o=4,6,8,10$, using empty squares for $u_o=4$, empty triangles for $u_o=8$, and the same symbols as in the main figure for the other cases. Curves in the inset are zooming in the range at which these curves change their curvature. 
Let us mention that curves overlap for values of $m \lesssim 1+ \log_2 (u_o+1)$, as can be predicted
from our calculations. In all the simulations, $N=1001$. We performed 50 runs, each one of $T=100000$ time steps discarding the first $50000$ steps for cases $u_o=0,2,4,6$, and $T=50000$ time steps discarding the first  $10000$ steps for the cases $u_o=8,10$. 
}
\label{fig9}
\end{figure}

In this section we present numerical results addressing the behaviour of the biased $MG$ studied in Subsection \ref{bisc}. We first plot $\sigma^2/N$ vs. $\alpha$ for different values of initial scores, always for $p=1/2$ as the probability of assigning the initial scores to the strategies. The main plot in Figure
\ref{fig9} shows these results for $u_o=2,6,10$. Initial scores decrease the waste of resources only during the crowd dynamics. 
 In the inset of the same figure, we present detailed results for $u_o=4,6,8,10$, zooming on the range at which these curves change their curvature. 
 Let us mention that, from equation (\ref{laultima}), it is easy to see that for values of $u_o$ greater than $n_o-1$, the expression for
 $\mathcal N_{d_{\tilde o}}^E-\mathcal N_{d_{\sim \tilde o}}^E$ {\it becomes independent of $u_o$} (indeed we get 
 $\mathcal N_{d_{\tilde o}}^E-\mathcal N_{d_{\sim \tilde o}}^E= \mathcal L^2/2^{2n_o} 2{2n_o-1\choose n_o-1}$).
 In particular, taking the ``typical'' $n_o=\mathcal H/2$ for the $MG_{rand}$, we see that 
   if $u_o \gtrsim \mathcal H/2-1$, then 
 a curve of $\sigma^2/N$ vs. $\alpha$ for a given   value of $u_o$, overlaps with any other curve given by a greater values of $u_o$ in the region of $m \lesssim 1+ \log_2 (u_o+1)$ (see Figure 9, e.g., if $u_o=6$ then the condition is met for $m \lesssim 3.8$).  
 
\begin{figure}[h]
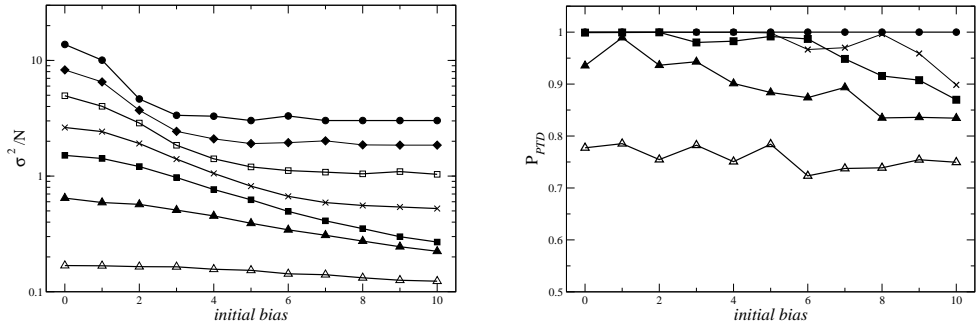

%\centering \includegraphics[width=5cm]{fig10.eps}
\centering\includegraphics[width=6cm]{fig10.eps} \qquad \includegraphics[width=6cm]{fig10c.eps}
\caption{Left: \emph{a)} $\sigma^2/N$ vs. initial scores at a fixed value of $\alpha$ (we take $N=1001$, and $m=2\cdots 8$), with filled circles ($m=2$), diamonds ($m=3$), empty squares ($m=4$), $X$ symbols ($m=5$), filled squares ($m=6$), filled triangles ($m=7$) and empty triangles ($m=8$). Right: \emph{b)} $P_{PTD}$ in even  occurrences of the states as a function of initial scores at a fixed value of $\alpha$. In the simulations we have computed $P_{PTD}$ as we did in Figure \ref{fig1} for the cases $N=1001$ and $m=2,5,6, 7$ and $8$. Symbols are the same as in figure \emph{a)}. In both figures, we performed 50 runs, each one of $T=100000$ time steps discarding the first $50000$ steps for cases $u_o=0,2,4,6$, and $T=50000$ time steps discarding the first  $10000$ steps for the cases $u_o=1,3,5,7,8$.
\label{fig10}
}
\end{figure}

 Figure \ref{fig10} a) and Figure \ref{fig10} b) show $\sigma^2/N$ and the probability $P_{PTD}$, respectively, vs. the initial scores for several values of  $\alpha$. In all cases, $N=1001$ is considered. Let us notice that PTD is valid in the biased model for $m=2$ ($\alpha \sim 0.004$) in all the range of initial biased values. If $m=5$ ($\alpha \sim 0.032$) then PTD is broken for $u_o=6$,  if $m=6$ ($\alpha \sim 0.064$) then PTD is broken for $u_o=3$, and if $m=7$ ($\alpha\sim 0.128$) or $m=8$ ($\alpha\sim 0.256$) then PTD is broken for all values of biased scores starting from $u_o=1$, as in the case of unbiased $MG$ (although for $m=7$, $P_{PTD}$ is very close to 1 when $u_o=1$).

It is worth to remark that since the maximum
of $2p_b(1-p_b)$ is reached for $p_b=1/2$ (in fact, this is the case in our simulations), we see that at most half of the agents (in average) are allowed to have different initial  point scores
for their strategies. Remarkably, as it is easily deduced from our calculations the reduced variance $\sigma^2/N$  behaves as $\alpha^{-1}$, as $u_o \to \infty$. To be more precise, for 
each $\alpha\gtrsim 0$ it is enough to take $u_o \gtrsim N\alpha=\mathcal H$ in order to obtain an analogous of equation (\ref{invariantemgrand}), that reads:
\begin{equation}
	 \sigma^2/N \sim 1/4+\frac{1}{16}\frac{N}{\pi\mathcal H},
	\label{cotainferior}
\end{equation}
in contrast with the case of uniform bias \cite{libro-MG,heimel exact}, for which  
$\sigma^2/N\sim \alpha$ when  $u_o\to \infty$.

\medskip
\medskip
\medskip

\medskip

\section{Conclusions}
\label{conclusions}

In this paper we showed that the $FSMG$ can be useful  to understand certain features of the $MG$ in the symmetric phase. The $FSMG$ is a maximal instance of the $MG$, where a single
copy of every potential agent takes part of the game.  Due to this fact, several symmetries can be exploited, allowing 
us to obtain analytical solutions for the $FSMG$. These theoretical  results were used to compute approximated values of the key variable $\sigma^2/N$ for the standard $MG$,
as well as for other versions based on different updating rules that can be found  in the literature. 
It is also shown that our technique allows to handle certain cases of strategies with biased initial scores.
We were able to show that the $FSMG$ enjoys the 
strict period two dynamics, a fact that led us to a simple way of computing 
the probability of breaking
 the period two dynamics for the first time in a given realization of the $MG$.  We are convinced that the $FSMG$ and the
framework presented can be useful, in the symmetric phase, for other variants of the $MG$. 

\section*{Acknowledgments}
The authors wish to thank the anonymous referees for several valuable suggestions, and particularly for the comments about the possibility of computing the biased case that led us to the results given in Section 3.3.

G. Acosta and I. Caridi are members of the  CONICET, Argentina.
This work has been partially supported by ANPCyT under grant BID PICT 2007-910.

%\section*{References}

%% References


\end{document}